\documentclass[aps,prl,preprint,groupedaddress,notitlepage]{revtex4}
\usepackage{graphicx}

\preprint{NuFact15 - Rio de Janeiro, Brazil - August 2015}

\begin{document}

\title{A non-conventional neutrino beamline  for the measurement of the 
electron neutrino cross section}
\thanks{\it Presented at NuFact15, 10-15 Aug 2015, Rio de Janeiro, 
Brazil [C15-08-10.2]}

\author{A. Berra, M. Prest}
\affiliation{Dep. of Physics, Univ. of Insubria and INFN, Como and Sezione di Milano-Bicocca, Milano, Italy}
\author{S. Cecchini, F. Cindolo, G. Mandrioli, N. Mauri, L. Patrizii, M. Pozzato, G. Sirri}
\affiliation{INFN, Sezione di Bologna, Bologna, Italy}
\author{C. Jollet, A. Meregaglia}
\affiliation{IPHC, Universit\'e de Strasbourg, CNRS/IN2P3, Strasbourg, France}
\author{L. Ludovici}
\affiliation{INFN, Sezione di Roma, Rome, Italy}
\author{A. Longhin, A. Paoloni, F. Pupilli, L. Votano}
\affiliation{Lab. Nazionali di Frascati dell'INFN, Frascati (RM), Italy}
\author{L. Pasqualini}
\affiliation{Dep. of Physics, Univ. of Bologna and INFN, Sezione di Bologna, Bologna, Italy}
\author{F. Terranova}
\email[]{francesco.terranova@cern.ch}
\thanks{Speaker}
\affiliation{Dep. of Physics, Univ. of Milano-Bicocca and INFN, Sezione di Milano-Bicocca, 
Milano, Italy}
\author{E. Vallazza}
\affiliation{INFN, Sezione di Trieste, Trieste, Italy}


\date{\today}

\begin{abstract}
  Absolute neutrino cross section measurements at the GeV scale are
  ultimately limited by the knowledge of the initial $\nu$ flux. In order
  to evade such limitation and reach the accuracy that is
  needed for precision oscillation physics ($\sim 1$\%), substantial
  advances in flux measurement techniques are requested. We discuss
  here the possibility of instrumenting the decay tunnel to identify
  large-angle positrons and monitor $\nu_e$ production from $K^+
  \rightarrow e^+ \nu_e \pi^0$ decays. This non conventional technique
  opens up opportunities to measure the $\nu_e$ CC cross section at the per cent level
  in the energy range of interest for DUNE/HK. We discuss the progress in the
  simulation of the facility (beamline and instrumentation) and the
  ongoing R\&D.
\end{abstract}


\maketitle

\section{Introduction}

A precise measurement of neutrino interaction cross sections will play
a key role in the next generation of oscillation physics experiments
and will impact significantly on the CPV and mass hierarchy (MH) reach
of long baseline facilities (see e.g.~\cite{nuint_2015}). This is particularly
evident for $\nu_e$ cross sections since $\nu_\mu \rightarrow \nu_e$
transitions (and their CP conjugate) represent the main observable to
measure the CP phase and determine the sign of $\Delta m^2_{31}$ (MH).

In the last ten years, an intense experimental programme has been
pursued, employing both the near detectors of running long-baseline
experiments and dedicated cross section
experiments~\cite{nuint_2015}. This programme already provided a
wealth of new data on absolute and differential cross section both
with inclusive (CC and NC) and exclusive final states
identification. These data challenge current theoretical
interpretations of neutrino interaction on nuclei at the GeV scale and
boosted the development of several new models and a systematic
comparison of existing approaches~\cite{Benhar:2015wva}.

Modern cross section experiments are swiftly reaching the intrinsic
limitations of conventional neutrino beams. In these beamlines, both
the $\nu_e$ and $\nu_\mu$ flux is inferred by a full simulation of
meson production and transport from the target down to the beam dump and
is validated by external data (hadro-production data, online
monitoring of the protons on target and muon current after the beam
dump).  Employing dedicated hadro-production experiments (replica
targets) the uncertainty on the neutrino flux can be reduced to
$\sim$10\% and additional improvements in the $5-10$\% scale are still
possible~\cite{bravar_nufact2015}.

On the other hand, reaching the per cent scale requires a change of
paradigm in the techniques employed to determine the neutrino flux
similar to the one recently proposed by the nuSTORM
Collaboration~\cite{Adey:2013pio}.

A technique with a similar aim as nuSTORM and specifically focused on
$\nu_e$ cross sections has been considered in~\cite{Longhin:2014yta}:
a beamline with focused and sign-selected secondaries at 8.5~GeV that
are transported down to an instrumented decay tunnel where electron
neutrinos are produced by the three body decay of $K^+$ ($K_{e3}$,
i.e. $K^+ \rightarrow e^+ \nu_e \pi^0$).  Inside this non conventional
decay tunnel, large angle positrons are identified by purely
calorimetric techniques. The mean energy and momentum bite ($\pm
20$\%) of the transfer line is optimized to enhance the $\nu_e$
components from $K_{e3}$ and suppress to a negligible level the
$\nu_e$ contamination from muon decays.  This beamline provides an
intense source of electron neutrinos for the study of $\nu_e$~CC
interactions. It exploits an observable (the positron rate) that can
be directly linked to the rate of $\nu_e$ at the far detector through
the three body kinematics of $K_{e3}$. The positron rate in the decay
tunnel allows for the {\it direct} monitoring of the $\nu$ rate at
source and provides a per cent measurement of the flux.

The proposal put forward in~\cite{Longhin:2014yta} must be validated through a
dedicated R\&D. The most relevant items are the design and
optimization of the beamline, the choice of the technology for the
positron monitoring and the evaluation of the systematic budget.  In
this talk, we report on the progress of such R\&D, the results
achieved in the last few months and the plans for the future.

\section{Yield at the target and transfer line}

\begin{figure}
\includegraphics[width=\textwidth]{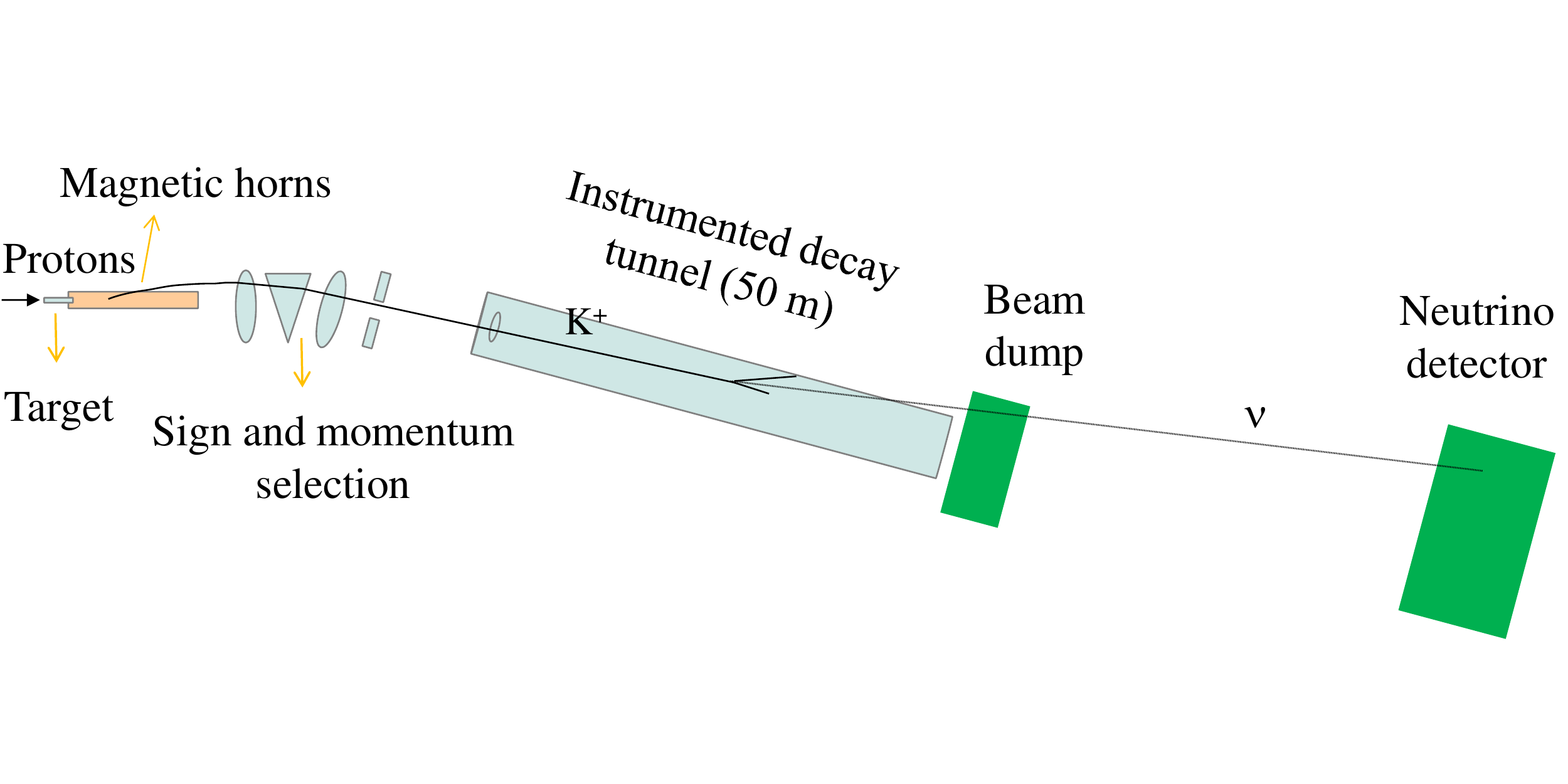} 
\caption{Layout of the facility (not to scale).}
\label{fig:schematics_nufact15}
\end{figure}

High precision $\nu_e$ cross section measurements based on $K_{e3}$
decays can be performed employing conventional beamlines with primary
protons impinging on a target, producing secondary hadrons which are
captured, sign selected and transported further down to the
instrumented decay tunnel (see Fig.~\ref{fig:schematics_nufact15}).

Secondary meson yields for this facility were evaluated with FLUKA
2011~\cite{Battistoni:2007zzb} to simulate primary proton interactions
on a 110~cm long (about 2.6 interaction lengths) cylindrical beryllium
target of 3~mm diameter. Graphite and INCONEL targets are being
simulated, too.  For the momentum range and transfer line acceptance
of interest for this study, the secondary yields at the target grow
linearly with the primary proton energy. The yields have been computed
simulating proton energies of relevance for the JPARC Proton
Synchrotron (30 GeV), the Fermilab Main Injector (120 GeV) and the CERN
SPS (450~GeV).

An end-to-end simulation of the focusing and transfer line is not
available yet and will be the subject of upcoming R\&D
work. Following~\cite{Longhin:2014yta}, fluxes at the entrance of the decay
tunnel are estimated considering the phase space $xx^{'}$, $yy^{'}$ of
pions and kaons in a momentum bite of 8.5~GeV/c~$\pm$~20\% at 5~cm
downstream the 110~cm long target.  All secondaries within an
emittance $\epsilon_{xx^{'}}=\epsilon_{yy^{'}}=0.15$~mm~rad are
focused assuming a typical horn focusing efficiency of
85\%~~\cite{Adey:2013pio}. 

Table~\ref{tab:yield} summarizes the results.  The second
and third columns show the pions and kaons per proton-on-target (PoT)
transported at the entrance of the decay tunnel. The fourth column
shows the number of PoT in a single extraction spill to obtain
$10^{10}$ pions per spill. The last column shows the number of
integrated protons on target that are needed to collect $10^4$ $\nu_e$ charged current
events on a 500 tons neutrino detector.  These proton fluxes are well
within the reach of the above-mentioned accelerators both in terms of
integrated PoT (from $5 \times 10^{20}$ at 30~GeV to $5 \times
10^{19}$ at 450~GeV) and protons per spill ($2.5 \times 10^{12}$ to $3
\times 10^{11}$).

\begin{table}
\begin{center}
\begin{tabular}{ccccc}
\hline
\hline

$E_p$ & $\pi^+$/PoT & $K^+$/PoT & PoT for a $10^{10}$ $\pi^+$  & PoT for $10^{4}$ $\nu_e$~CC \\
 (GeV)       &($10^{-3}$) &($10^{-3}$)&   spill  ($10^{12}$)           &  ($10^{20}$)        \\
\hline   
30 [JPARC]      &	 4.0  &  0.39     &     2.5                  &    5.0              \\
120 [Fermilab]    &	16.6  &  1.69     &     0.60                 &    1.16            \\ 
450 [CERN]    &	33.5  &  3.73     &     0.30                 &    0.52            \\
\hline   
\hline
\end{tabular}
\caption{Pion and kaon yields at
(8.5$\pm$1.7)~GeV/c. The rightmost column is evaluated assuming a 500
  ton neutrino detector located 50~m after the beam dump.}
\label{tab:yield}
\end{center}
\end{table}

\section{Proton extraction scheme}

The results of the previous Section combined with the maximum particle
rate sustainable by the instrumentation of the decay tunnel (see
below) fix the main constraint on the length of the proton spill
extracted from the accelerator. For a maximum particle rate of
500~kHz/cm$^2$, this constraint corresponds to an upper limit to the
average number of PoT per second:
\begin{equation}
\mathrm{PoT/s} \ < \ 1.5 \times 10^{14}
\label{eq:constraint}
\end{equation}
For instance, assuming 450~GeV protons extracted from the SPS (third
line of Table~\ref{tab:yield}), a 2~ms (10~ms) spill requires less than
$3 \times 10^{11}$ PoT/spill ($1.5 \times 10^{12}$ PoT/spill). This
operation mode is unpractical for high energy machines (e.g. the SPS)
where the number of protons circulating in the lattice exceeds
$10^{13}$ but the repetition rate is O(0.1)~Hz. These machines must
hence resort to (resonant) slow extraction modes. Two options are
currently under investigation:
\begin{itemize}
\item {\bf Slow extraction modes}: a 1~s slow extraction mode similar
  to the one devised for SHiP at the CERN-SPS~\cite{SHIP_arduini}. It
  is the classical solution envisaged for the ``tagged neutrino
  beams''~\cite{denisov,Ludovici:2010ci} and it fulfills the
  constraint of Eq.~(\ref{eq:constraint}) even in the occurrence of complete
  depletion of the protons accumulated in the lattice ($4.5 \times
  10^{13}$ for the CERN-SPS).  It comes with two significant drawbacks:
  it prevents the use of magnetic horns and challenges the cosmic
  background reduction of the neutrino detector. Still, due to the
  relatively low flux needed for cross section measurements compared
  with standard oscillation experiments, static focusing systems based
  on FODO/FFAG~\cite{nupil} represent a viable option for this facility.
\item {\bf Multiple slow resonant extractions}: Slow extractions of
  limited duration (10~ms, a few thousands turns) repeated frequently
  ($\sim 10$~Hz) can be envisaged to deplete the lattice at the end of
  the acceleration phase. For instance, in the standard operation mode
  of the CERN-SPS~\cite{SHIP_arduini}, particles are extracted during
  a flat top of 4.8~s inside the 15~s full acceleration cycle
  (super-cycle). This mode corresponds to a 30\% duty cycle. Assuming
  $4.5 \times 10^{13}$ accumulated protons in the lattice, full
  depletion is achieved with 30 extractions of 10~ms ($1.5 \times
  10^{12}$ PoT per extraction) repeated every 160~ms. The feasibility
  of this kind of schemes for the particular case of the CERN-SPS is
  under investigation.
\end{itemize}

In both cases, assuming full depletion mode (i.e. the accelerator
running in dedicated mode for the neutrino experiment) the integrated
exposure requested in Table~\ref{tab:yield} to perform the cross
section measurement is reached in $\sim$200 days ($\sim$ 1 year
considering a standard 200 days/y effective livetime).

\section{Instrumented decay tunnel}

In conventional low energy neutrino beams, the decay tunnel is located
just after the horn and therefore accepts neutral
and wrong sign particles, together with high energy protons. Doses and
rates are therefore not suitable for additional instrumentation.  In
the facility considered here (Fig.~\ref{fig:schematics_nufact15}) the
decay tunnel is located at the end of the transfer line, while neutrals
and protons are dumped before the bending dipoles. In addition, the
positrons produced by kaon decays have a polar angle that is much
larger than muons from $\pi^+ \rightarrow \mu^+ \nu_\mu$
decays. Additional instrumentation can hence be located just in the
outer radius of the tunnel. Undecayed pions, transported protons and
muons from pion decay will reach the beam dump without intercepting
the outer walls of the decay tunnel and, hence, will not contribute to
the rates. This is assured by the above constraint on the emittance: if
the entrance windows of the secondaries in the tunnel and the spread
in polar angle is smaller than the muon production angle from pion
decay (4~mrad for 8.5~GeV pions), all particles (but decayed kaons)
will reach the beam dump without additional focusing units inside the
tunnel. In~\cite{Longhin:2014yta}, a 50~m long tunnel with 40~cm inner radius
and a $\pm5$~cm entrance windows with polar angles smaller than 3~mrad
has been considered. The precise values are under evaluation in the
framework of the end-to-end simulation of the transfer line.

\begin{figure}
\includegraphics[width=0.8\textwidth]{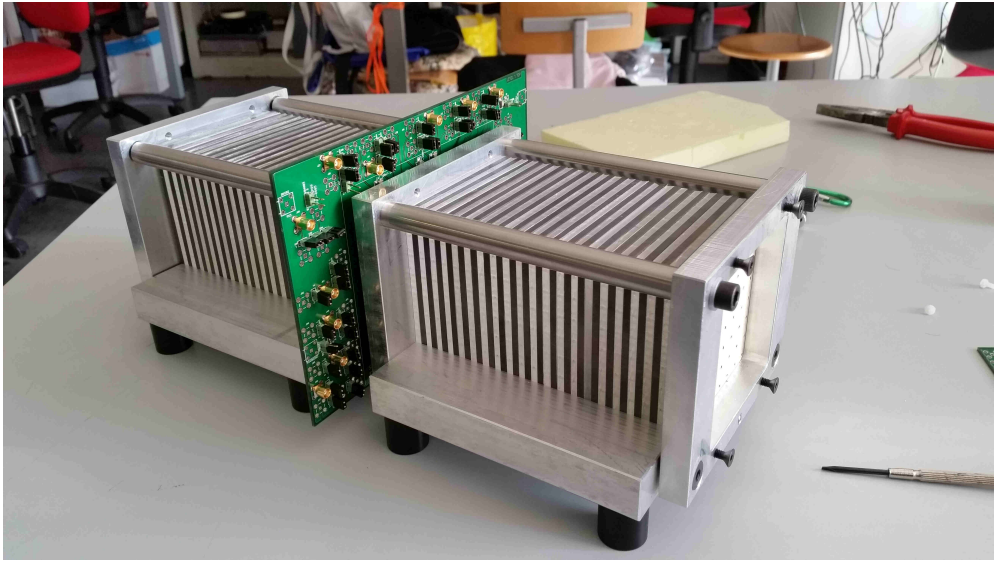} 
\caption{Test prototype for the light readout system of the
  calorimeter based on SiPM embedded in the shashlik module.}
\label{fig:proto_photo}
\end{figure}

The most critical issue is the identification of the detector
technology that can be used to instrument (a fraction of) the
evacuated ($<$1 mbar) decay tunnel. As for the general study performed
in~\cite{Longhin:2014yta}, the detector must be able to stand a
maximum rate of 500~kHz/cm$^2$ and provide charged pions/positron
misidentification and photon veto at few percent level. Radiation
hardness must be assured at the level of $>1.3$~kGy.  Shashlik
calorimeters with fast fiber readout and longitudinal segmentation
(sampling every 4~$X_0$) complemented by a plastic scintillator photon
veto offer a compact and cost effective solution, which has already
been proved to be radiation hard at the $>$5~kGy
level~\cite{poliakov_ship}. Both ionizing and non ionizing (neutron)
doses are low enough to allow for the use of solid state photosensors
(SiPM) embedded inside the module of the calorimeters. Each SiPM reads
separately a WLS fiber of the module and the outputs of multiple
SiPM's are summed up.  Full simulation of this setup is in progress
and, for modules of $3\times 3 \times 10$~cm$^2$ size (sum of 9 SiPM),
preliminary results confirm the positron identification capability
estimated for a generic calorimeter in~\cite{Longhin:2014yta}. The
embedding of the SiPM inside the modules to achieve longitudinal
segmentation without loss of compactness and with negligible dead
zones has been tested in summer 2015 with an early prototype
(Fig.~\ref{fig:proto_photo}) at CERN PS. The test demonstrated that
nuclear counter effects are negligible and the embedding does not
introduce significant deterioration of the energy response with
respect to standard fiber bundling~\cite{berra}.

\section{Event rates and systematics contributions}

The high momentum  (8.5~GeV) secondaries selected in the transfer line produce
a neutrino beam at the end of the decay tunnel that is enriched in
$\nu_e$ from kaon decays and depleted in $\nu_e$ from muon decay in flight (DIF).
For the parameters of~\cite{Longhin:2014yta}, the $\nu_e/\nu_\mu$ flux ratio at the neutrino
detector is independent of the proton energy and it is:
$$
\frac{\Phi_{\nu_e}}{\Phi_{\nu_\mu}} = 1.8 \ \% \ (\nu_e \ \mathrm{from} \ K_{e3}) 
\ \ ; \ \  
\frac{\Phi_{\nu_e}}{\Phi_{\nu_\mu}} = 0.06 \ \% \ (\nu_e \ \mathrm{from \ DIF}). 
$$
The mean energy of the neutrinos interacting at the far detector
($\nu_e$ CC events) is 3~GeV with a FWHM of $\sim$3.5~GeV. This region
covers the entire range of interest for the next generation long
baseline experiments. Unlike conventional neutrino beams, a facility
that is able to monitor the positron production at the decay tunnel
can provide a flux estimate that does not depend on prior information
on the proton intensity and secondary yields. A summary of the most
relevant contributions is given in Tab.~\ref{tab:systematics}. Current
activities focus on the evaluation of the sub-dominant contributions
due to the instrumentation response in the decay tunnel.

\begin{table}
\begin{center}
\begin{tabular}{lllp{6cm}}
\hline
\hline

Uncertainty & Conv. & This  &  \\
\hline   
kaon (pion) production yield     &	X  &       &                   \\ 
kaon/pion ratio     &	X  &       &                   \\ 
protons-on-target     &	X  &       &                \\ 
statistical error on monitored $e^+$ &  & X & $<$0.1\% \\
geometrical efficiency     &	X  & X      &  survey ($<$0.5\%)              \\ 
3 body kinem. and $K^+$ mass     &	X  & X      &  $<0.1$\%            \\ 
phase space at tunnel entrance &  & X & measured on-site \\ 
BR of $K_{e3}$ & X &  &  \\ 
e/$\pi$/$\gamma$ separation &  & X & measured with test beams and onsite with control samples \\
calorimeter response stability &  & X & on-site monitoring and calibration \\ 
residual gas in beampipe &  & X & negligible at 0.1 mbar\\ 
\hline   
\hline
\end{tabular}
\caption{Main contributions to flux uncertainty for conventional (``Conv.'') neutrino beams 
and for this facility (``This''). ``X'' indicates whether the contribution is relevant or is by-passed by the
monitoring of the positrons.}
\label{tab:systematics}
\end{center}
\end{table}

\section{Conclusions}

The knowledge of the flux at source in conventional neutrino beams
dominates the precision of neutrino cross section measurements in
short baseline experiments. In order to reach a per cent accuracy, a
breakthrough in the experimental techniques employed to estimate the
flux is needed. The technique we are investigating is particularly
well suited for the measurement of the $\nu_e$ cross section - a key
ingredient to establish CP violation in the leptonic sector - and it
is based on the monitoring of large angle positrons originating from
$K^+ \rightarrow e^+ \nu_e \pi^0$. We discussed the most relevant
technical challenges and ongoing R\&D both for the design of the
beamline and for the instrumentation of the decay tunnel.  In
particular, we identified a specific detector option based on shashlik
calorimetry that is suitable for the instrumentation of the decay
tunnel and fulfills the requirements of PID capability, pile-up
mitigation and radiation hardness.



\bibliography{terranova}

\end{document}